\documentclass[preprint2]{aastex}     
\catcode`\@=11
\def\gsim{\ifmmode{\mathrel{\mathpalette\@versim>}}
    \else{$\mathrel{\mathpalette\@versim>}$}\fi}
\def\lsim{\ifmmode{\mathrel{\mathpalette\@versim<}}
    \else{$\mathrel{\mathpalette\@versim<}$}\fi}
\def\@versim#1#2{\lower 2.9truept \vbox{\baselineskip 0pt \lineskip
    0.5truept \ialign{$\m@th#1\hfil##\hfil$\crcr#2\crcr\sim\crcr}}}
\catcode`\@=12

\def\msol{{\rm M}_\odot}
\def\lsol{{\rm L}_\odot}
\def\mpc{\rm Mpc}
\def\kpc{{\rm kpc}}
\def\rhovir{\rho_{\Delta}}
\def\r200{r_{\Delta}}
\def\rvir{r_{\rm vir}}
\def\sigvir{\sigma_{\rm vir}}
\def\re{R_{\rm e}}
\def\rh{R_{\rm h}}
\def\sigmah{\sigma_{\rm h}}
\def\mlgal{\Upsilon_{\rm gal}}

\begin{document}
\title{THE SCALING RELATIONS OF GALAXY CLUSTERS AND THEIR DARK MATTER HALOS} 
\author{B. Lanzoni$^1$, L. Ciotti$^{2,3}$, A. Cappi$^1$, G. Tormen$^4$, 
        G. Zamorani$^1$} 
\affil{$^1$INAF -- Osservatorio Astronomico di Bologna, via Ranzani 1,
       40127, Bologna, Italy \\ 
       $^2$ Dipartimento di Astronomia, Universit\`a di Bologna, 
       via Ranzani 1, 40127 Bologna, Italy \\
       $^3$ Scuola Normale Superiore, Piazza dei Cavalieri 7, 56126 , Pisa,
       Italy \\
       $^4$ Dipartimento di Astronomia, Universit\`a di Padova, 
       vicolo dell'Osservatorio 5, 35122 Padova, Italy }

\date{September 19, 2003; accepted}

\begin{abstract}
Like early-type galaxies, also nearby galaxy clusters define a Fundamental
Plane, a luminosity-radius, and a luminosity-velocity dispersion relations,
whose physical origin is still unclear. By means of high resolution N--body
simulations of massive dark matter halos in a $\Lambda$CDM cosmology, we find
that scaling relations similar to those observed for galaxy clusters are
already defined by their dark matter hosts. The slopes however are not the
same, and among the various possibilities in principle able to bring the
simulated and the observed scaling relations in mutual agreement, we show
that the preferred solution is a luminosity dependent mass-to-light ratio
($M/L\propto L^{\sim 0.3}$), that well corresponds to what inferred
observationally.  We then show that at galactic scales there is a conflict
between the cosmological predictions of structure formation, the observed
trend of the mass-to-light ratio in ellipticals, and the slope of their
luminosity-velocity dispersion relation (that significantly differs from the
analogous one followed by clusters). The conclusion is that the scaling laws
of elliptical galaxies might be the combined result of the cosmological
collapse of density fluctuations at the epoch when galactic scales became
non-linear, plus important modifications afterward due to early-time
dissipative merging.  Finally, we briefly discuss the possible evolution of
the cluster scaling relations with redshift.
\end{abstract}

\section{Introduction}
\label{sec:intro}
Early-type galaxies are known to follow well defined scaling relations
involving their main observational properties, i.e, the luminosity
$L$, effective radius $\re$, and velocity dispersion $\sigma$: in
particular we recall here the Faber-Jackson (hereafter FJ; Faber \&
Jackson 1976), the Kormendy (Kormendy 1977), and the Fundamental Plane
(hereafter FP; Djorgovski \& Davis 1987; Dressler et al. 1987)
relations.  Within the limitations of a poorer statistics, analogous
relations have been found to hold also for galaxy clusters (Schaeffer
et al. 1993, hereafter S93; Adami et al. 1998a, hereafter A98; for
3-parameters scaling relations involving X--ray observables, see Annis
1994; Fujita \& Takahara 1999a; Fritsch \& Burchert 1999; Miller,
Melott \& Gorman 1999).  Besides their potential importance as
distance indicators, these scaling relations are also useful to get
insights on the structure and, possibly, on the formation and
evolutionary processes of galaxies and galaxy clusters.

Well defined scaling relations, that recall the observed ones, are indeed
expected on the basis of the simplest model for the formation of structures
in an expanding Universe, namely the gravitational collapse of density
fluctuations in an otherwise homogeneous distribution of collisionless dark
matter (DM).  In fact, the spherical top-hat model (Gunn \& Gott 1972)
predicts that, at any given epoch, all the existing DM halos have just
collapsed and virialized, i.e., $M = \rvir\,\sigvir^2/G$ (where $\rvir\equiv
-U/G\,M^2$, $\sigvir^2\equiv 2T/M$, $U$ and $T$ being the potential and
kinetic energies of a halo of mass $M$, respectively).  In addition, all the
halos are characterized by a constant mean density $\rhovir$, given by the
critical density of the Universe at that redshift times a factor $\Delta$
depending on $z$ and on the given cosmology (see, e.g., Peebles 1980; Eke,
Cole \& Frenk 1996). For simplicity, we call $\r200$ the radius of the sphere
containing such a mean density, so that $M\propto\r200^3$. In general,
$\rvir\ne\r200$, but, if for a family of density distributions
$\rvir/\r200\simeq const$, then the virial theorem can be rewritten as $M
\propto\r200\,\sigvir^2$, and, together with $M\propto \r200^3$, it brings to
$M\propto\sigvir^3$, thus providing three relations that closely resemble the
observed ones.  Note that these expectations involve the {\it global
three-dimensional} properties of DM halos, while the quantities entering the
observed scaling relations are {\it projected} on the plane of the sky.
However, if DM halos are structurally homologous\footnote{Strictly speaking,
DM halos are only {\it weakly} homologous systems, since the low mass halos
are systematically more concentrated than the more massive ones (for a
definition of weak homology, see Bertin, Ciotti \& del Principe 2002).}
systems, as found in cosmological simulations (Navarro, Frenk \& White 1997,
hereafter NFW), and are characterized by similar velocity dispersion profiles
(e.g., Cole \& Lacey 1996), their projected properties are also expected to
follow well defined scaling relations (with some scatter due to departures
from perfect homology and sphericity).

Of course, the simple considerations above are not sufficient to
account for the {\it observed} scaling relations of galaxy clusters,
at least for two reasons. The first is that a given potential well (as
the one associated with the cluster DM distribution) can be filled, in
principle, by very different distributions of ``tracers'' (such as the
galaxies in the clusters, from which the scaling relations are
derived).  This means that the very existence of the cluster FP
implies a remarkable regularity in their formation processes: galaxies
must have formed or ``fallen'' in all clusters in a similar way.  The
second reason is that any trend of the cluster mass-to-light ratio
(necessary to transform masses, involved in the theoretical relations,
into luminosities\footnote{We recall here that the luminosity of a
cluster refers to the sum of the luminosities of all its constituent
galaxies, i.e., a sum over the cluster luminosity function.}, entering
the observed ones) must be taken into account for a proper
interpretation of the observed scaling relations.

A distinct but strongly related question about the origin and the
meaning of the scaling laws naturally arises when applying the
predictions of the cosmology also at galactic scales: in fact, while
scale-invariant relations are predicted, different slopes of the FJ
relation are observed for galaxies and for galaxy clusters (see
Section \ref{sec:obs}, and Girardi et al. 2000).  This suggests that
different processes have been at work in setting or modifying the
correlations at the two mass scales.  The theoretical implications of
the scaling laws for elliptical galaxies (Es) have been intensively
explored (see, e.g., Bertin et al. 2002 and references therein), and
several works have been devoted to their study within the framework of
the dissipationless merging scenario in Newtonian dynamics (e.g.,
Capelato, de Carvalho \& Carlberg 1995; Nipoti, Londrillo \& Ciotti
2003a, 2003b, hereafter NLC03ab; Evstigneeva, Reshetnikov \& Sotnikova
2002; Dantas et al. 2003; Gonz\'alez-Garc\'{\i}a \& van Albada 2003).
In particular, if the initial total energy of the system is
non-negative, the merger products are found to follow the observed
edge-on FP (with the important exception of merging dominated by
accretion of small galaxies), but they badly fail at reproducing the
FJ and the Kormendy relations, in accordance with elementary
predictions based on energy conservation and on the virial theorem
(NLC03ab). It has also been shown (Ciotti \& van Albada 2001) that gas
free mergers cannot account at the same time for the FP and for the
$M_{\rm BH}$-$\sigma$ relation (that link the black hole mass and the
stellar velocity dispersion of the host spheroid, Ferrarese \& Merrit
2000; Gebhardt et al. 2000); the importance of gas dissipation in the
formation of elliptical galaxies is also apparent from the observed
color--magnitude and ${\rm Mg}_2$-$\sigma$ relations (e.g., Saglia et
al. 2000; Bernardi et al. 2003a).  Much less effort has been devoted
to the theoretical study of the FP of galaxy clusters: beside a work
on the effects of dissipationless merging (Pentericci, Ciotti \&
Renzini 1996), the other few theoretical studies mainly focus on the
possible relation between the FP properties and the cluster age, the
number of substructures, and the underlying cosmology (Fujita \&
Takahara 1999b; Beisbart, Valdarnini \& Buchert 2001). From the
comparison between the FP of galaxies and of galaxy clusters in the
$k$-space, Burstein et al. (1997) derived support to the idea that, at
variance with groups and clusters, gas dissipation must have had an
important role on the formation and evolution of galaxies.

In order to get a more complete view of the problems depicted above,
we use high-resolution N-body simulations to study the scaling
relations of very massive DM halos, that are thought to host the
present day galaxy clusters.  The aim is to verify whether these
relations are similar or not to the observed ones, and determine the
assumptions required to make them in mutual agreement.  The results
thus obtained and the empirical evidence of different slopes of the
scaling relations at galactic and cluster scales are then discussed.
The paper is organized as follows: in Section \ref{sec:obs} we present
the scaling relations observed for nearby galaxy clusters and those
followed by early-type galaxies. In Section \ref{sec:DM} we describe
the high-resolution resimulation technique employed to build the
sample of massive DM halos used for our analysis, and we derive their
scaling relations. In Sections \ref{sec:simu_obs} and \ref{sec:cl_gal}
we discuss under which hypothesis these scaling relations can be
translated into the observed ones, focusing in particular on the
astrophysical implications of the differences at galactic and cluster
scales. The main results are summarized and discussed in Section
\ref{sec:disc}.

\section{Scaling relations of clusters and early-type galaxies}
\label{sec:obs}
From the observational point of view only two works (namely, S93 and
A98) report on the FP of galaxy clusters, i.e., a relation among their
optical luminosity, scale radius, and velocity dispersion.  The two
groups agree about the existence of a tight and well defined FP, even
if quantitative differences in the numerical values of its
coefficients are found, that can be traced back to the different
choice of the radial variable. In fact, A98 use 4 density profiles
(King, Hubble, NFW, and de Vaucouleurs) to describe the distribution
of cluster galaxies, concluding that the best fit is provided by the
first two models: consistently, they adopt the clusters core radii to
obtain the FP.  S93 instead use in their analysis the cluster
half-light projected radii (the standard choice in the vast majority
of the FP studies).  For this reason we choose to compare our results
to those presented by S93: we note however that the FP coefficients
derived by A98 when using the projected half-light radii derived from
the de Vaucouleurs model agree within the errors with those reported
by S93.

\begin{figure}
\plotone{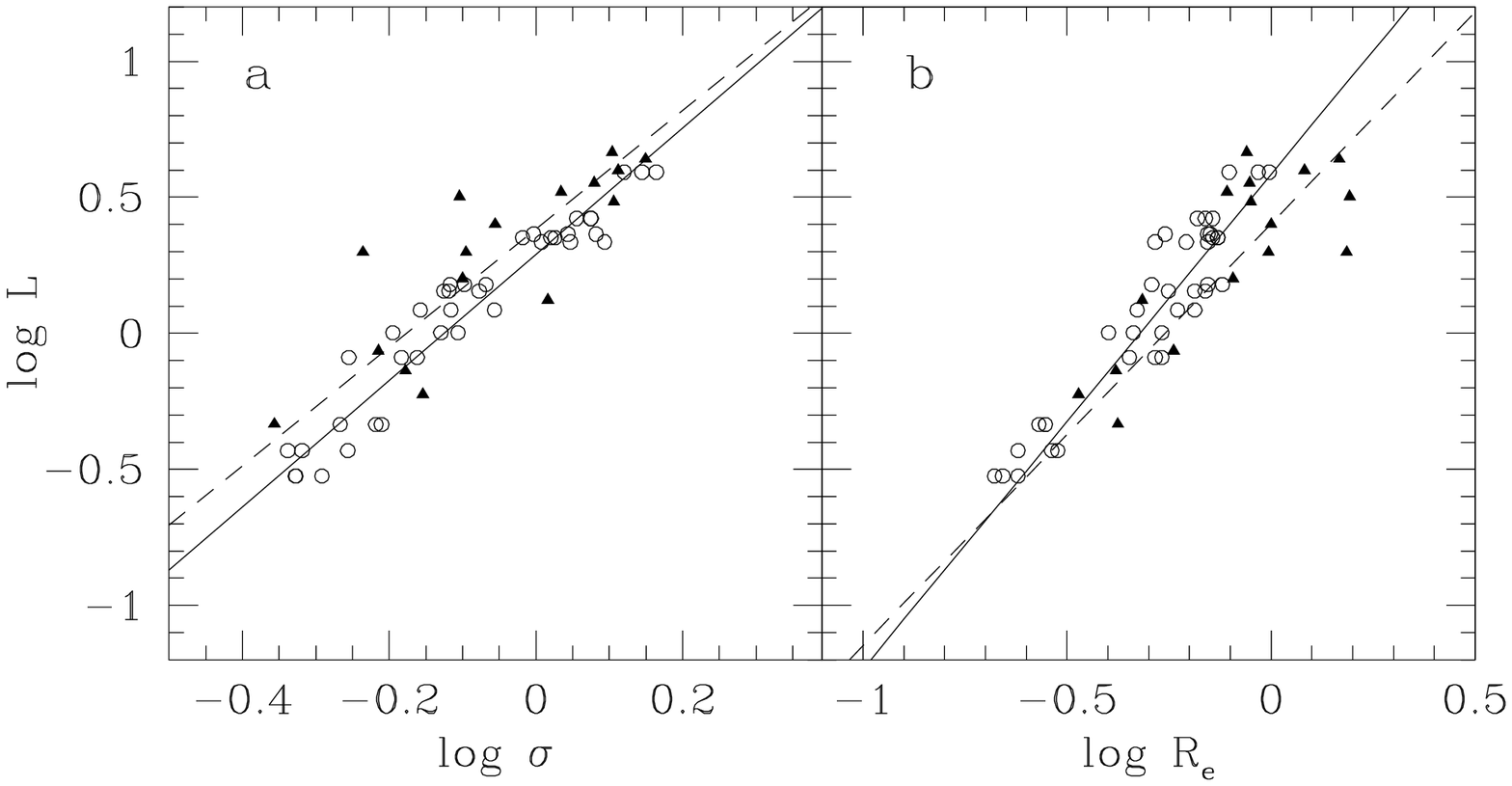}
\caption{{\it Panel a:} FJ relation for the observed clusters (filled
triangles) and the DM halos when equation (\ref{eq:ml}) is used for the
mass-to-light ratio (empty circles).  The corresponding best fit relations
are $L\propto \sigma^{2.18}$ (dashed line) and $L\propto \sigma^{2.32}$
(solid line), respectively.  {\it Panel b:} The Kormendy relation for the
same data as in Panel a. The best fit relations are $L\propto \re^{1.55}$ for
the data (dashed line), and $L\propto \re^{1.82}$ for the DM halos (solid
line).  Each DM halo is represented by 3 empty circles corresponding to the 3
line-of-sight directions. Luminosities are normalized to $10^{12}\,\lsol/h$,
velocity dispersions are in 1000 km/s, and radii in Mpc/$h$.
\label{fig:fjk}}
\end{figure}

\begin{figure}
\plotone{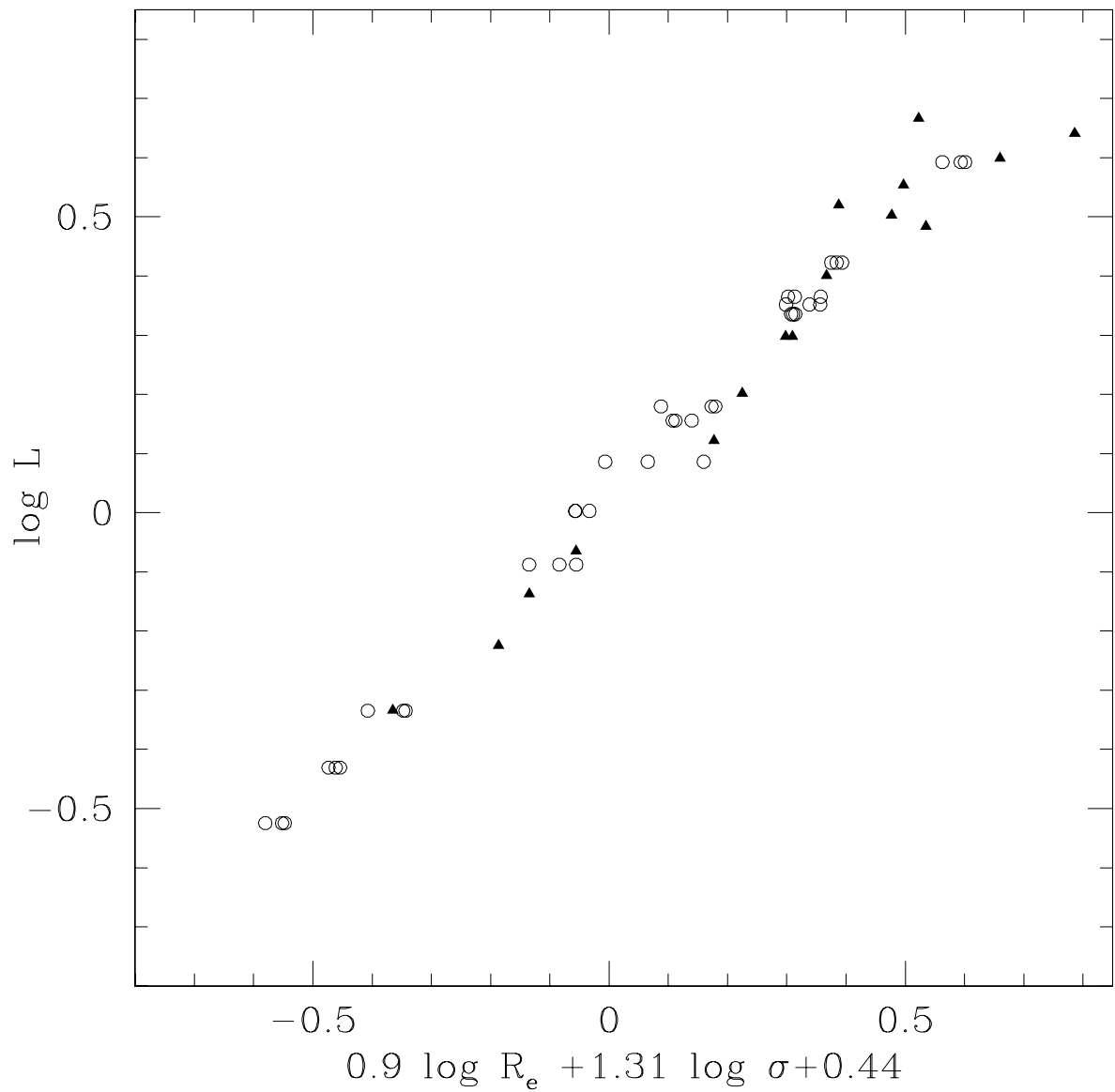}
\caption{The FP of observed clusters (filled triangles) and DM halos when
equation (\ref{eq:ml}) is used for the mass-to-light ratio (empty circles).
\label{fig:fpl}}
\end{figure}

For their sample of 16 galaxy clusters at $z\le 0.2$, S93 used the
photometric parameters $L$ and $\re$ in the V band (quoted by West,
Oemler \& Dekel, 1989), and the velocity dispersion $\sigma$ (given by
Struble \& Rood, 1991), and they derived not only the FP, but also a
FJ-like and a Kormendy-like relations.  However, instead of reporting
the S93 scaling laws, that have been obtained by means of least-square
fits to the data, here we re-derive them by minimizing the distance of
the residuals perpendicular to a straight line (for the FJ and the
Kormendy relations) or to a plane (for the FP).  Results are anyway
consistent with those of S93. The FJ and Kormendy relations we obtain
are:
\begin{equation}
L\propto \sigma^{2.18\pm0.52}, 
\label{eq:fj}
\end{equation}
in good agreement also with the results of Girardi et al. (2000), and
\begin{equation}
L\propto \re^{1.55\pm0.19}, 
\label{eq:kor}
\end{equation}
where $L$ is given in $10^{12}L_\odot/h^2$, $\sigma$ in 1000 km/s,
$\re$ in Mpc$/h$ ($H_0 = 100\,h$ km s$^{-1}$ Mpc$^{-1}$, and $h=1$),
and the errors on the exponents take into account also the
observational uncertainties.  As can be seen from Fig.\ref{fig:fjk},
where equations (\ref{eq:fj}) and (\ref{eq:kor}) are plotted together
with the S93 data, the two relations above describe real scalings
among the cluster properties, even if their scatter is quite large:
without taking into account the observational errors, the $rms$
dispersion of the data around the best-fit lines is 0.19 in both
cases.  Similarly to what happens for galaxies, a considerable
improvement is achieved when combining all the three observables
together in a FP relation, that we have derived by performing a
Principal Component Analysis (PCA; see e.g. Murtagh \& Heck 1987) on
the data sample, thus obtaining the new orthogonal variables $p_i$
defined by:
\begin{equation}
p_i\equiv\alpha_i \log\re + \beta_i\log L + \gamma_i \log\sigma, ~~ i=1,2,3.
\label{eq:pi}
\end{equation}
The numerical values of the coefficients $\alpha_i$, $\beta_i$, and
$\gamma_i$ are listed in Table \ref{tab:pca}, while the resulting
distribution of the observed clusters in the ($p_1,p_3$) and ($p_1,
p_2$) spaces are shown in Fig.\ref{fig:fp_pi}, the former providing an
exact edge-on view of the FP and making apparent its small thickness.
A more intuitive representation of a (nearly) edge-on view of the FP
can be obtained by solving equation (\ref{eq:pi}) for $i=3$ and using
$p_3\simeq const$:
\begin{equation}
L\propto\re^{0.9\pm 0.15} \,\sigma^{1.31\pm 0.22}.
\label{eq:fp}
\end{equation}
This relation is shown in Fig.\ref{fig:fpl}, and is characterized
by an $rms$ dispersion of $\sim 0.07$.

How do cluster scaling relations compare with the analogous relations
followed by early-type galaxies?  For what concerns the FP, the
agreement with that of galaxies is remarkable. For example, the FP of
elliptical galaxies in the B band is given by $L \propto \re^{\sim
0.8}\, \sigma^{\sim 1.3}$ (e.g., Dressler et al. 1987; J{\o}rgensen,
Franx \& Kj{\ae}rgaard 1996; Scodeggio et al. 1998; Bernardi et
al. 2003c). As well known, the exponents of the galaxy FP depend on
the adopted photometric band (see, e.g., Scodeggio et al. 1998; Treu
2001), but the differences from the values reported above become
significant only when using the K-band (Pahre, Djorgovski \& de
Carvalho 1998). Thus, the agreement between the cluster and the galaxy
FP can be regarded as robust. The situation is similar for the
Kormendy relation: in fact, $L\propto \re^{\sim 1.7\pm0.07}$ has been
reported for ellipticals in the B band (Davies et al. 1983; see also
Schade, Barrientos \& Lopez-Cruz 1997; Bernardi et
al. 2003b). However, this agreement should be regarded as less robust
than that of the FP, because largely different values of the Kormendy
slope for various selections of the data sample have been reported in
the case of galaxies (Ziegler et al. 1999). In any case, the FJ
relation of galaxies is different from that the clusters FJ: in fact,
$L\propto\sigma^4$ for (luminous) Es (Faber \& Jackson 1976; Forbes \&
Ponman 1999; Bernardi et al. 2003b), while for clusters the reported
slope is around 2 (a value of $\sim 1.58$ is also found by A98).  The
additional facts that the slope of the galaxy FJ seems to drop below 3
for ellipticals with $\sigma<170$ km/s (see, e.g., Davies et
al. 1983), and that the exponent of the $M_{\rm BH}$-$\sigma$ relation
(Ferrarese \& Merrit 2000; Gebhardt et al. 2000) is remarkably similar
to that of the FJ will be discussed in Section \ref{sec:cl_gal}.

\begin{deluxetable}{rrrr}
\tablewidth{0pt}
\footnotesize
\tablecaption{PCA parameters for the observed clusters.}
\tablehead{ \colhead{} & $\alpha_i$ & $\beta_i$ & $\gamma_i$ } 
\startdata
$p_1$ & -0.67 & -0.51  & -0.94     \\
$p_2$ &  0.88 & -0.005 & -1.22     \\
$p_3$ &  0.55 & -0.61  &  0.80     \\
\enddata
\tablecomments{$p_i$ is defined in equation(\ref{eq:pi}).}
\label{tab:pca}
\end{deluxetable}

\section{DM halos scaling relations}
\label{sec:DM}
\subsection{The simulations}
\label{sec:simu}
To investigate whether the dark matter hosts of galaxy clusters, as
obtained by numerical simulations, do define scaling relations similar
to the observed ones, a large enough sample of very massive DM halos
is needed.  Therefore, we employed dissipationless simulation with
$512^3$ particles of $6.86\times 10^{10}\msol/h$ mass each, where the
volume of the Universe is sufficiently large for this purpose: the box
side is $479\,h^{-1}\mpc$ comoving, with $h=0.7$ (see Yoshida, Sheth
\& Diaferio 2001).  The adopted cosmological model is a $\Lambda$CDM
Universe with $\Omega_{\rm m}=0.3$, $\Omega_\Lambda=0.7$ (e.g.,
Ostriker \& Steinhardt 1995), spectral shape $\Gamma =0.21$, and
normalization to the local cluster abundance, $\sigma_8=0.9$.  From
this simulation, we have randomly selected a sub-sample of 13 halos at
$z=0$, with masses between $10^{14}\msol/h$ and $2.3\times
10^{15}\msol/h$. They span a variety of shapes, from nearly round to
more elongated.  The richness of their environment also changes from
case to case, with the less isolated halos usually surrounded by
pronounced filamentary structures, containing massive neighbors (up to
$20\%$ of the selected halo in mass).

Given the mass resolution of the simulation, less than 1500 particles
compose a halo of $10^{14}\msol/h$ and, due to discreteness effects,
its properties defining the FP relation cannot be accurately
determined.  We have therefore {\it resimulated} at higher resolution
the halos in our sample by means of the technique introduced in
Tormen, Bouchet \& White (1997): here we recall only its relevant
aspects (for more details see Lanzoni, Cappi \& Ciotti 2003).  The
first step is to select in a given cosmological simulation the halo
one wants to ``zoom in''.  Then, the region defined by all the
particles composing the selected halo and its immediate surroundings
is detected in the initial conditions of the parent simulation, and
the number of particles within it is increased by the factor needed to
attain the suited mass resolution. Such a region is therefore called
``the high resolution region'' (HRR).  Since the mean inter-particle
separation within the HRR is smaller than in the parent simulation,
the corresponding high-frequency modes of the primordial fluctuation
spectrum are added to those on larger scales originally used in the
parent simulation, and the overall displacement field is also modified
consequently.  At the same time, the distribution of surrounding
particles is smoothed by means of a spherical grid, whose spacing
increases with the distance from the center: in such a way, the
original surrounding particles are replaced by a smaller number of
\emph{macroparticles}, whose mass grows with the distance from the
HRR.  Thanks to this method, even if the number of particles in the
HRR is increased, the total number of particles to be evolved in the
simulation remains small enough to require reasonable computational
costs, while the tidal field that the overall particle distribution
exerts on the HRR remains very close to the original one.  For the new
initial configuration thus produced, vacuum boundary conditions are
adopted, i.e., we assume a vanishing density fluctuation field outside
the spherical distribution of particles with diameter equal to the
original box size $L$.  A new N-body simulation is then run starting
from these new initial conditions, and allows to re-obtain the
selected halo at the required resolution.

We have applied this technique to the selected 13 massive DM halos.  For 8 of
them the resolution has been increased by a factor $\sim 33$, by means of
high-resolution particles of mass $\sim 2.07\times 10^9\msol/h$ each, while a
further increase of a factor of 2 has been adopted for the 5 intermediate
mass halos (the particle mass is $10^9\msol/h$ in this case).  The
gravitational softening used for the high-resolution region is $\epsilon =
5\,\kpc/h$ (roughly Plummer equivalent), corresponding to about $0.2\%$ and
$0.5\%$ of the virial radius of the most and least massive halos,
respectively. This scale length represents the spatial resolution of the
resimulations, to be compared with that of $30\,\kpc/h$ of the original one.
To run the resimulations, the parallel dissipationless tree-code GADGET
(Springel, Yoshida \& White 2001) has been used. At $z=0$ the DM halos have
been selected by means of a spherical overdensity criterium (Lacey \& Cole
1994; Tormen et al. 1997), i.e., they are defined as spheres centered on
maximum density peaks in the particle distribution, and with mean density
equal to the virial density $\rhovir$ predicted by the spherical top-hat
model for the adopted $\Lambda$CDM cosmology ($\rhovir \simeq 97 \,\rho_{\rm
crit}$ at $z=0$). The corresponding mass, linear scales and velocity
dispersions are listed in Table \ref{tab:simu}.

\begin{deluxetable}{rrccrcccrrr}
\tabletypesize{\scriptsize} 
\tablewidth{0pt} 
\tablecaption{Properties of the DM halos at $z=0$.} 
\tablehead{ \colhead{Name} & \colhead{$M$} & \colhead{$\r200$} & 
\colhead{$\rvir/\r200$} & \colhead{$\sigvir$} & \colhead{$R_{{\rm h},x}$} &
\colhead{$R_{{\rm h},y}$} & \colhead{$R_{{\rm h},z}$} &
\colhead{$\sigma_{{\rm h},x}$} &  \colhead{$\sigma_{{\rm h},y}$} &
\colhead{$\sigma_{{\rm h},z}$} \\ \colhead{} & \colhead{$10^{14} \msol/h$} &
\colhead{Mpc$/h$} & \colhead{} & \colhead{km/s} & \colhead{Mpc$/h$} &
\colhead{Mpc$/h$} & \colhead{Mpc$/h$} & \colhead{km/s} & \colhead{km/s} &
\colhead{km/s}} 
\startdata
 1542 &  0.82 & 0.99 & 0.77 &  760 & 0.22 & 0.24 & 0.21 &  471 &  470 &  511 \\
 3344 &  1.09 & 0.99 & 0.83 &  815 & 0.30 & 0.24 & 0.29 &  459 &  554 &  480 \\
  914 &  1.45 & 1.09 & 0.75 &  967 & 0.27 & 0.28 & 0.28 &  616 &  605 &  541 \\
 4478 &  2.92 & 1.37 & 1.02 & 1075 & 0.54 & 0.52 & 0.45 &  556 &  656 &  689 \\
 1777 &  3.83 & 1.50 & 0.94 & 1177 & 0.40 & 0.54 & 0.46 &  783 &  638 &  742 \\
  564 &  4.91 & 1.63 & 1.01 & 1271 & 0.65 & 0.59 & 0.47 &  696 &  878 &  766 \\
  689 &  6.08 & 1.75 & 1.01 & 1354 & 0.56 & 0.65 & 0.69 &  837 &  749 &  761 \\
  245 &  6.50 & 1.79 & 1.03 & 1332 & 0.51 & 0.70 & 0.76 &  855 &  799 &  764 \\
   51 & 10.78 & 2.12 & 0.84 & 1794 & 0.62 & 0.70 & 0.52 & 1114 & 1018 & 1241 \\
  696 & 11.37 & 2.16 & 0.98 & 1669 & 0.74 & 0.72 & 0.74 & 1062 & 1048 &  959 \\
   72 & 11.77 & 2.18 & 0.90 & 1777 & 0.70 & 0.71 & 0.55 & 1105 &  993 & 1207 \\
    1 & 13.99 & 2.31 & 0.88 & 1870 & 0.66 & 0.72 & 0.69 & 1187 & 1136 & 1189 \\
    8 & 23.42 & 2.75 & 0.92 & 2262 & 0.79 & 0.93 & 0.99 & 1459 & 1395 & 1319 \\
\enddata
\label{tab:simu}
\tablecomments{$M$ is the DM halo
mass, $\r200$ and $\rvir$ are the truncation and virial radius,
respectively, $\sigvir$ is the virial velocity dispersion; the
projected half-mass radii and the corresponding line-of-sight velocity
dispersions are given for three orthogonal directions. Note that
$\rh/\r200$ is in the range $\sim 0.2$--0.4.}
\end{deluxetable}

\subsection{Scaling relations for the DM halos}
\label{sec:DMsr}
According to the selection method described in the previous section,
the DM halos are all characterized by the same $\rhovir$ and should
also be nearly virialized systems.  The first condition (and thus the
relation $M\propto \r200^3$) is satisfied by construction while the
second can be easily verified.  In fact, we find that all the selected
halos follow the relations $M\propto \r200\,\sigvir^2$ with a $rms$
scatter of 0.03 only, and $M\propto\sigvir^{3.1}$, with $rms\simeq
0.05$. In Table \ref{tab:simu} we list the ratio $\rvir/\r200$ for
each halo.

However, {\it projected} quantities are involved in the observations
and the first step of our analysis is the determination of which (if
any) scaling relations are satisfied by the DM halos when projected.
We have therefore constructed the projected radial profiles of the
selected halos by counting the DM particles within concentric shells
around the center of mass for three arbitrary orthogonal directions
($x$, $y$, and $z$), and we defined $\rh$ as the projected radius of
the circle containing half of the total number of particles. Then, the
velocity dispersion $\sigmah$ has been computed from the line-of-sight
(barycentric) velocity of all the particles within $\rh$.  Since the
DM halos, as well as real clusters, are not spherical, such a
procedure gives different values of $\rh$ and $\sigmah$ for the three
line-of-sights (the maximum variations however never exceed 33\% and
21\% for the two quantities, respectively; see Table \ref{tab:simu})
and we decided to build our data sample by considering all the three
projections for each halo.  With the projected properties $\rh$ and
$\sigmah$ now available, we have determined the best fit relations
between $M$ and $\rh$, and between $M$ and $\sigmah$ by minimizing the
distance of the residuals perpendicular to a straight line, and thus
obtaining the DM analogues of the observed FJ and Kormendy relations:
\begin{equation}
M\propto\sigmah^{3.02\pm 0.15}, 
\label{eq:DM_fj}
\end{equation}
and 
\begin{equation}
M\propto \rh^{2.36\pm 0.14}, 
\label{eq:DM_kor}
\end{equation}
with $rms \simeq 0.12$ and 0.15, respectively.  With a PCA of these data we
determined the relation analogous to equation (\ref{eq:fp}):
\begin{equation}
M\propto \rh^{1.1\pm 0.05}\sigmah^{1.73\pm 0.04}, 
\label{eq:DM_fp}
\end{equation}
with $rms=0.04$.  Compared to those among the virial properties, these
relations have larger scatters, as expected.  The FJ and FP have slopes
similar to those obtained for the virial quantities, while the $M$-$\rh$
relation appears to be significantly flatter, as a consequence of the
different density concentration of low and high mass halos. Note that the NFW
concentration parameter varies only weakly in the mass range covered by the
halos in our sample (e.g., Eke, Navarro \& Steinmetz 2001; Bullock et
al. 2001). This is consistent with our results: in fact, from equation
(\ref{eq:DM_kor}) and $M\propto \r200^3$, we find that $\rh/\r200 \propto
M^{\sim 0.09}$.

We stress that while scaling relations between $M$, $\r200$ (or $\rvir$) and
$\sigvir$ were expected, a tight correlation between projected properties is
a less trivial result: in fact, structural and dynamical non-homology can, in
principle, produce significantly different effective radii and projected
velocity dispersion profiles for systems characterized by identical $M$,
$\rvir$, and $\sigvir$.  It is also known that weak homology, coupled with
the virial theorem, does indeed produce well defined scaling laws (see, e.g.,
Bertin et al. 2002). Therefore, the scaling relations presented in equations
(\ref{eq:DM_fj})--(\ref{eq:DM_fp}) are a first interesting result of our
study.  The difference between the values of the exponents appearing in
equations (\ref{eq:DM_fj})--(\ref{eq:DM_fp}) and those in the virial
relations is the direct evidence of weak homology of the halos: in fact,
strong homology would result in exactly the same exponents, while a strong
non-homology would disrupt any correlation.  We checked this further, and we
found that the DM halos in our sample still show well defined scaling
relations (although with different exponents) when considering the projected
radius encircling any fixed fraction of the total mass, and the corresponding
projected velocity dispersion within it (see Table \ref{tab:shells}).  Note
that this finding is in agreement with the results already pointed out by
several groups, namely the fact that DM halos obtained with numerical N-body
simulations in standard cosmologies are characterized by significant
structural and dynamical weak homology (e.g., Cole \& Lacey 1996; NFW;
Subramanian, Cen \& Ostriker 2000, and references therein).

\begin{deluxetable}{rrrrrrrrrrrr}
\tabletypesize{\scriptsize} 
\tablewidth{0pt} 
\tablecaption{Exponents and $rms$ scatter of the scaling relations
$M\propto \sigma_{fr}^a$, $M\propto R_{fr}^b$,  and $M\propto R_{fr}^c
\, \sigma_{fr}^d$.} 
\tablehead{ \colhead{$fr$} & \colhead{$a$} & \colhead{$rms$} & \colhead{$b$}
& \colhead{$rms$} & \colhead{$c$} & \colhead{$d$} & \colhead{$rms$}}  
\startdata
0.8 & 3.01 & 0.12 & 2.67 & 0.10 & 1.47 & 1.43 & 0.04 \\
0.5 & 3.02 & 0.12 & 2.36 & 0.15 & 1.10 & 1.73 & 0.04 \\
0.3 & 3.10 & 0.13 & 2.29 & 0.17 & 0.97 & 1.95 & 0.04 \\
0.1 & 3.12 & 0.13 & 2.15 & 0.19 & 0.88 & 2.02 & 0.03 \\
\enddata
\label{tab:shells}
\tablecomments{$M$ is halo total mass, $R_{fr}$ is the projected
radius containing the mass $fr\times M$, and $\sigma_{fr}$ is the mean
projected velocity dispersion within $R_{fr}$: $R_{0.5}\equiv \rh$.}
\end{deluxetable}

\section{From the simulated to the observed scaling relations}
\label{sec:simu_obs}
Comparison of equations (\ref{eq:DM_fj}), (\ref{eq:DM_kor}),
(\ref{eq:DM_fp}) and (\ref{eq:fj}), (\ref{eq:kor}), (\ref{eq:fp}),
reveals that the FJ, Kormendy and FP relations of simulated DM halos
are characterized by different slopes with respect to those derived
observationally.  What kind of regular and systematic trend with
cluster mass of the galaxy properties and distribution are implied by
these differences? In order to answer this question, we define the
dimensionless quantities $\Upsilon\equiv M/L$, ${\cal
R}\equiv\rh/\re$, and ${\cal S}\equiv\sigmah/\sigma$. Focusing first
on the edge-on FP, from equations (\ref{eq:fp}) and (\ref{eq:DM_fp})
we obtain:
\begin{equation}
\frac{\Upsilon}{{\cal{R}}^{1.1}\,{\cal{S}}^{1.73}} \propto
\re^{0.2}\sigma^{0.42}.
\label{eq:cfr_fp}
\end{equation}
Thus, in order to transform the DM halos FP into the observed FP, the product
$\Upsilon {\cal{R}}^{-1.1} {\cal{S}}^{-1.73}$ must systematically increase as
$\re^{0.2}\sigma^{0.42}$, which in turn, again from equation (\ref{eq:fp}),
is approximately proportional to $L^{0.3}$.  In principle, $\Upsilon$,
${\cal{R}}$, and ${\cal{S}}$ could all vary in a {\it combined} and {\it
regular} way from cluster to cluster, so that equation (\ref{eq:cfr_fp}) is
satisfied. Of course, given the small scatter around the best fit relation
(\ref{eq:fp}), this kind of solution requires a remarkable fine tuning of the
variations of the three parameters. Alternatively, it is possible that only
one of the three parameters varies significantly, while the other two are
approximately constant.  This situation is analogous to that faced in the
studies of the physical origin of the FP tilt of elliptical galaxies, where
the so called ``orthogonal exploration of the parameter space'' is often
adopted (see, e.g., Renzini \& Ciotti 1993; Ciotti 1997) . In this approach
all but one of the available model parameters are fixed to constant values,
and the goal is to determine what kind of variation of the ``free'' parameter
is necessary to reproduce the observed FP tilt.  Several quantitative results
have been derived in this framework (see, e.g., Ciotti, Lanzoni \& Renzini
1996; Ciotti \& Lanzoni 1997, and references therein), even though, by
construction, it cannot provide the most general solution to the problem, and
the choice of the specific parameter responsible for the tilt is somewhat
arbitrary (see Bertin et al. 2002; Lanzoni \& Ciotti 2003).  In the present
context some of this arbitrariness can be removed: in fact, here we assume
that 1) the DM distribution in real clusters is described by the simulated DM
halos, and that galaxies are merely dynamical {\it tracers} of the total
potential well, 2) in addition to the edge-on FP, we also consider the
constraints imposed by the FJ and the Kormendy relations.  These two points
will {\it allow to use the orthogonal exploration approach for determining
what is the most plausible origin of the tilt between the simulated and the
observed cluster FP}.

In order to make the DM halos FP reproduce the observed one within the
framework of the orthogonal exploration approach, we have three
different possibilities, each corresponding to the choice of
$\Upsilon$, ${\cal{R}}$, or ${\cal{S}}$ as the key parameter, while
keeping constant the remaining two in the l.h.s. of equation
(\ref{eq:cfr_fp}).  Note that the two choices based on variations of
${\cal{R}}$ or ${\cal{S}}$ should be interpreted from an astrophysical
point of view as systematic differences in the way galaxies populate
the cluster DM potential well as a function of the cluster mass.
However, the orthogonal analysis of the FJ and the Kormendy relations
strongly argue against these two solutions, since from equations
(\ref{eq:fj}), (\ref{eq:kor}), (\ref{eq:DM_fj}), and (\ref{eq:DM_kor})
one obtains
\begin{equation}
\frac{\Upsilon}{{\cal{S}}^{3.02}} \propto \sigma^{0.84}, 
\label{eq:cfr_fj}
\end{equation}
and
\begin{equation}
\frac{\Upsilon}{{\cal{R}}^{2.36}} \propto\re^{0.81}. 
\label{eq:cfr_kor}
\end{equation}
Thus, it is apparent that any attempt to reproduce equation
(\ref{eq:cfr_fp}) by a variation of ${\cal R}$ (or ${\cal S}$) alone
will fail at reproducing the FJ (or the Kormendy) relation: in fact,
the only parameter appearing in all the equations (\ref{eq:cfr_fp}),
(\ref{eq:cfr_fj}), and (\ref{eq:cfr_kor}) is the mass-to-light
ratio\footnote{Note that the constraints imposed by the FJ and the
Kormendy relations should not be considered redundant with respect to
those imposed by the edge-on FP: in fact, these two relations, albeit
with a large scatter, describe how galaxies are distributed on the
face-on FP.} $\Upsilon$.

\begin{figure}
\plotone{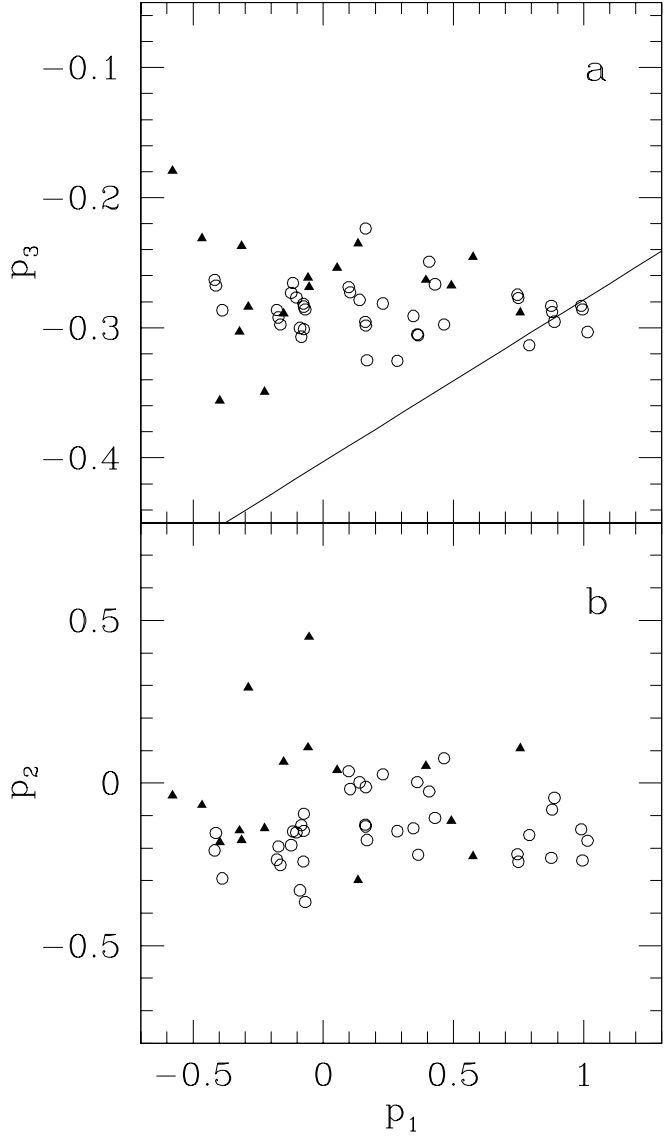} \figcaption{The edge-on (upper panel) and face-on
(lower panel) views of the FP for the observed clusters (filled triangles)
and the DM halos when equation (\ref{eq:ml}) is used for the mass-to-light
ratio (empty circles). The solid line would be the best fit of the DM halos
when assuming a constant mass-to-light ratio $\Upsilon =
280\,h\,\msol/\lsol$. Note that mass increases for decreasing $p_1$.
\label{fig:fp_pi}}
\end{figure}

Therefore, while a {\it purely structural} (${\cal R}$) and a {\it
purely dynamical} (${\cal S}$) origin of the tilt between the DM halos
FP and the clusters FP seem to be both ruled out by the reasons above,
a systematically varying mass-to-light ratio, for ${\cal R}$ and
${\cal S}$ constant, could in principle account for all the three
considered scaling relations.  In particular, from equation
(\ref{eq:cfr_fp}), $\Upsilon\propto\,L^{\alpha}$ with $\alpha\sim
0.3$. Guided by this indication, we tried to superimpose the points
corresponding to the simulated DM halos to the sample of observed
clusters by using $\Upsilon\propto M^\beta$, and we found that if
\begin{equation}
\Upsilon = 280\, h\,\left(\frac{M}{10^{14}\,\msol/h}\right)^{0.23}
\,\frac{\msol}{\lsol},
\label{eq:ml}
\end{equation}
{\it the edge-on FP of DM halos is practically indistinguishable from that of
real clusters} (see Fig.\ref{fig:fp_pi}a and Fig.\ref{fig:fpl}): the value of
$\alpha$ derived from this assumption is $\alpha=\beta/(1-\beta)\simeq 0.3$.
Note that such a trend of $\Upsilon$ is in agreement not only with the
expectations of equation (\ref{eq:cfr_fp}), but also with what inferred
observationally in the B-band (S93; Girardi et al. 2002; Bahcall \& Comeford
2002; but see Bahcall, Lubin \& Dorman 1995; Bahcall et al. 2000; and
Kochanek et al. 2003 for claims of a constant mass-to-light ratio at large
scales), and with what inferred from the comparison between the observed
B-band luminosity function of virialized systems and the halo mass function
predicted in CDM cosmogonies (Marinoni \& Hudson 2002).  A remarkable
agreement is also found with the results of van den Bosch, Yang \& Mo (2003),
who, using a completely different approach, obtain $\Upsilon\propto M^{0.26}$
for their model A, which allows for a non-constant $\Upsilon$ at the cluster
scales (see their equation 15 and Table 2).  In addition, $\Upsilon\sim 280
\,h\,\msol/\lsol$ for $M\simeq 10^{14}\,\msol/h$ clusters, and $\Upsilon\sim
475 \,h\,\msol/\lsol$ for $M\simeq 10^{15}\,\msol/h$ clusters are values well
within the range of the various estimates for galaxy clusters (e.g., Adami et
al. 1998b; Mellier 1999; Wilson, Kaiser \& Luppino 2001; Girardi et al. 2000;
Girardi et al. 2002).  It is also remarkable (as not a necessary consequence)
that by adopting equation (\ref{eq:ml}) {\it also the face-on FP, and the FJ
and Kormendy relations are very well reproduced}, as apparent from
Fig.\ref{fig:fp_pi}b and Figs.\ref{fig:fjk}ab, respectively.

What could be the physical interpretation of the required trend of the
mass-to-light ratio with cluster mass?  We note that equation
(\ref{eq:ml}) can be formally rewritten as $\mlgal\times(M/M_{\rm
gal}) \propto M^\beta$, where M is the DM mass of the clusters ($\sim$
their total mass), $M_{\rm gal}$ is their total stellar content in
galaxies,
\begin{equation}
\mlgal\equiv\frac{\int{N_{\rm gal}(L)\, \Upsilon_*(L)\, L\,
                  dL}}{\int{N_{\rm gal}(L)\, L\,dL}}, 
\end{equation}
where $N_{\rm gal}(L)$ is the cluster luminosity function, and finally
$\Upsilon_*(L)$ is the mean stellar mass-to-light ratio of a galaxy of
total luminosity $L$.  Thus, the trend of $\Upsilon$ with $M$ could be
ascribed or to $\mlgal\propto M^\beta$ for $M/M_{\rm gal} = const$, or
to $M_{\rm gal}\propto M^{1-\beta}$ for $\mlgal = const$ (or, more
generally, to a combined effect of these two quantities).  Both
possible solutions have interesting astrophysical implications. For
example, a constant $\mlgal$ from cluster to cluster is obtained only
if all clusters have the same population of galaxies, i.e., if they
are characterized by a universal luminosity function (LF) and by a
similar morphological mix, so that the distribution of the stellar
mass-to-light ratios of their galaxies is also the same.  In such a
case, the required trend $M_{\rm gal}\propto M^{0.77}$ should be
entirely explained by a systematic increase with $M$ of the total
number of galaxies, in remarkable agreement with several studies of
the halo occupation numbers, that find $N_{\rm gal}\propto M^a$, with
$a$ in the range 0.7--0.9 (Peacock \& Smith 2000; Scranton 2002;
Berlind \& Weinberg 2002; Marinon \& Hudson 2002; van den Bosch et
al. 2003).  To study the ratio $M/M_{\rm gal}$, an estimate of the
stellar mass in galaxies can also be obtained from the observed total
luminosity in the near infrared (since Es are the dominant component
of the cluster population and their luminosity in the K band gives a
reasonable measure of their stellar mass, i.e., $M_{\rm gal}\propto
L_K$).  However, observational results are still uncertain and
controversial: a constant or weakly decreasing $M/L_K$ with $M$ is
reported by Kochaneck et al. (2003) for the 2MASS clusters, while an
increasing $M/L_K$ is claimed by Lin, Mohr \& Standford (2003) for the
data from the same survey. 

In any case, the universality of the cluster luminosity function is
still under debate since it appears to be appropriate in many cases,
but variations in some individual clusters have also been reported
(see, e.g., Yagi et al. 2002, De Propris et al. 2003, and Christlein
\& Zabludoff 2003 for detailed discussions and recent results).  In
particular, the LF of early-type (or quiescent) galaxies appears to
vary significantly among clusters (but see Andreon 1998): if the
richer clusters contain a proportionally larger fraction of elliptical
galaxies (Balogh et al. 2002), that are characterized by a higher
$\Upsilon_*$ compared to that of spirals, a cluster-dependent $\mlgal$
should be expected even in presence of a universal LF.

\section{Cluster vs. galaxy scaling relations}
\label{sec:cl_gal}
As discussed in Section \ref{sec:obs}, the cosmological collapse model
predicts the same scaling relations at all mass scales and,
remarkably, the edge-on FP and the (although more dispersed) Kormendy
relation of clusters and ellipticals are very similar. In particular,
the interpretation of the FP tilt of elliptical galaxies as the
combined results of the virial theorem and strong structural and
dynamical homology, implies $\Upsilon_*\propto L^{0.3}$ in the B band
(e.g., Faber et al. 1987; van Albada, Bertin \& Stiavelli 1995; Bertin
et al. 2002), in agreement with the result for clusters (equation
\ref{eq:ml}).  In Section \ref{sec:simu_obs} we showed that equation
(\ref{eq:DM_fj}) coupled with $\Upsilon\propto L^{0.3}$ does reproduce
the observed FJ relation at cluster scales, but clearly this cannot
work in the case of galaxies: in fact, the FJ relation of Es is
characterized by a significantly higher exponent ($\sim 4$) than that
of clusters ($\sim 2$). Indeed, $\Upsilon_*$ should actually {\it
decrease} for increasing galaxy luminosity in order to reproduce the
observed FJ, at variance with all the available indications (e.g.,
Faber et al. 1987; J{\o}rgensen et al. 1996; Scodeggio et al. 1998).

One is then forced to assume that 1) the relation $M\propto\sigvir^3$ does
not apply in the case of Es, perhaps due a failure of standard cosmology at
small scales, or 2) evolutionary processes have modified the ratio
$\sigma_*/\sigvir$, where $\sigma_*$ is the galaxy central velocity
dispersion (from which the FJ relation is derived).  For what concerns point
1), this might be another problem encountered by the current cosmological
paradigm at small scales, in addition to the well known over-prediction of
the number of DM satellites (the so called ``DM crisis''; see Moore et
al. 1999; but see also St\"ohr et al. 2002 and the recent results on the
``running spectral index'' obtained by Peiris et al. 2003 from the WMAP data)
and the cusp-core problem for low surface brightness and dwarf galaxies
(e.g., Swaters et al. 2003ab, and references therein). However, this point is
at present very speculative, and thus in the following we restrict to the
standard CDM paradigm, that predicts $M\propto\sigvir^3$ also at galactic
mass scales. Two physical processes (namely, gas dissipation and early-time
merging) certainly played a major role in galaxy evolution, thus supporting
point 2).  The effects of these two processes have been already discussed in
the context of $k$-space by Burstein et al. (1997): here we will focus on the
implications that can be derived from the simpler FJ relation, taking into
account the additional information from recent numerical simulations
(NLC03ab) and the constraints imposed on galaxy merging by the recently
discovered $M_{\rm BH}$-$\sigma$ relation (Ferrarese \& Merritt 2000;
Gebhardt et al. 2000), according to which $M_{\rm BH}\propto\sigma_*^{\sim
4}$, with very small dispersion.

We then adopt the point of view that at the epoch of the detachment from the
Hubble flow, also the seed galaxies were characterized by the ``universal''
scaling law $M\propto\sigvir^3$, and we discuss the possibility that galaxy
merging and gas dissipation originated the observed FJ of Es. We start by
considering the effect of dissipationless merging alone. From this point of
view, it is clear that merging played different roles in the evolutionary
history of clusters and galaxies.  In fact, while clusters can be thought as
formed from the {\it collapse} of density perturbations at scales that {\it
just became non-linear}, galaxies are presently in a highly non-linear regime
and the {\it merging} they suffered since their separation from the Hubble
expansion can no longer be interpreted in terms of the cosmological collapse
of density fluctuations.  The differences between these two dynamical
processes have important consequences for the present discussion. In fact, in
the {\it cosmological collapse} case the initial conditions correspond to
those of a ``cold'' system (i.e., $2\,T_i+U_i=V<0$), and so virialization
increases the virial velocity dispersion of the end-products as $T_f=T_i-V$:
the systematic increase of $\sigvir$ with $M$ in clusters is basically due to
this process. At highly non-linear scales (such as, for instance, in the case
of galaxies in the outskirts of clusters or groups) the situation is
considerably different. In fact, if merging occurs, it cannot be interpreted
as the collapse of a cold system, but, on the contrary, the initial
conditions of the merging pair are in general characterized by a null or a
positive $V$ (see, e.g., Binney \& Tremaine 1987, Chap.7): under these
conditions, the virial velocity dispersion of the remnant will not
increase. In fact, numerical simulations of one and two-component galaxy
models show that successive dissipationless merging at galactic scales, while
preserving the edge-on FP, does not reproduce the FJ relation, being the
end-products characterized by a too low (i.e., nearly constant) $\sigma_*$
compared to the expectations of the FJ (NLC03ab). In other words,
dissipationless merging at galactic scales in general will produce a relation
$M\propto\sigvir^\alpha$ with $\alpha>3$. Obviously, this could be an
interesting property in our context, since we are exactly looking for a
mechanism able to increase $\alpha$ from 3 to 4.  We also note that faint
elliptical galaxies do follow a FJ relation with a best-fit slope
significantly lower than 4 (Davies et al. 1983 report a value of $\sim 2.4$)
and more similar to the cosmological predictions: a simple interpretation of
this fact could be that small galaxies experimented less merging events than
giant Es, so that their scaling relations are more reminiscent of the
cosmological origin.  However, even if able to increase the exponent of the
$M$-$\sigma$ relation, several theoretical and empirical arguments clearly
indicate that purely dissipationless merging cannot be at the origin of the
spheroids. For example, why the exponent should increase to the observed
value is unclear, even though it has been claimed that the FJ of Es is the
result of cumulative effects of inelastic merging and passing encounters
taking place in cluster environment (Funato, Makino \& Ebisuzaki 1993; Funato
\& Makino 1999). These authors already discuss some weakness of the scenario
they suggest (for instance, the difficulty to account for the analogous
Tully-Fisher relation of spirals), and we add that it is not clear whether
also the other scaling laws are reproduced by their simulations, nor how the
FJ relation could be followed also by the field Es if resulting from
dynamical interactions in cluster environment.  Moreover, a further
difficulty is added by the recently discovered $M_{\rm BH}$-$\sigma$
relation: if the FJ is the cumulative result of random dynamical processes
that changed the galaxy velocity dispersion, how can the $M_{\rm
BH}$-$\sigma$ relation be so tight?  Finally, an even stronger empirical
argument argues against substantial dissipationless merging for Es: in fact,
if the merging end-products are forced to obey both the edge-on FP and the
$M_{\rm BH}$-$\sigma$ relation, then they are characterized by exceedingly
large effective radii with respect to real galaxies (Ciotti \& van Albada
2001).

Concerning the effect of gas dissipation on the predicted relations at
galactic and cluster scales, we note that an important difference already
resides in {\it what} is effectively observed in the two cases when deriving
the scaling laws: the galaxy distribution, tracing the gravitational
potential of the dark matter, in the case of clusters; the stellar population
within a fraction of the effective radius, where stars themselves are the
major contributors to the total gravitational field, in the case of Es.
Since the top-hat model properly describes the (dissipationless) collapse of
the DM component, it is not surprising that it cannot be accurately applied
for predicting the properties of the baryonic (dissipative) matter where it
is dominant.  In addition, gas dissipation is thought to be negligible for
the formation and evolution of clusters, while it empirically appears to be
increasingly important with mass for the galaxies, as mirrored by their
$Mg_2$-$\sigma$ and color-magnitude relations, and by the observed
metallicity gradients (e.g., Saglia et al. 2000; Bernardi et al. 1998;
Bernardi et al. 2003a).  As discussed by Ciotti \& van Albada (2001), its
effects also represent a possible solution to the exceedingly large effective
radii found for (dissipationless) merger products forced to follow both the
FP and the $M_{\rm BH}$-$\sigma$ relation.  Thus, from the considerations
above, gas dissipation is a good candidate to explain the difference between
cluster and galaxy FJs. However, we note that {\it baryonic dissipation
alone} cannot solve the discrepancy: in fact, its effect is to increase
$\sigma_*$, thus further decreasing the exponent in the galactic FJ relation.
This can be seen in a more quantitative way as follows. Given that galactic
DM halos follows $M\propto \sigvir^3$, then the FJ relation implies that
$(M/M_*)\,\Upsilon_*\,(\sigma_*/\sigvir)^3\propto \sigma_*^{-1}$: this means
that at fixed $M/M_*$ and $\Upsilon_*$, dissipation should have been more
important in low mass galaxies, contrarily to the existing observational
evidences.

In summary, the arguments presented above strongly suggest that merging {\it
and} gas dissipation played a fundamental role in the formation and evolution
of galaxies. In particular, given the competitive effect of these two
processes on the slope of the $M$-$\sigma$ relation, they appear to be {\it
both} necessary for modifying the $M\propto\sigvir^3$ into the observed FJ.
Coupled with the necessity of dissipative merging, the observational evidence
that the bulk of stars in Es, even at high $z$, are old puts a strong
constraint on the time when the mass of spheroids was assembled: say, $z\,
\gsim\, 2$. As already recognized by several authors, this is in agreement
with the available information about the star formation history of the
Universe and the redshift evolution of the quasar luminosity function (see,
e.g., Haehnelt \& Kauffmann 2000; Ciotti \& van Albada 2001; Burkert \& Silk
2001; Yu \& Tremaine 2002; Cavaliere \& Vittorini 2002; Haiman, Ciotti \&
Ostriker 2003).  A tight connection between the star formation and the quasar
activity in Es is further supported by the striking similarity of slopes
between the FJ and the $M_{\rm BH}$-$\sigma$ relations, thus suggesting a
common origin for these two empirical laws.

Based on these considerations, it seems likely that the scaling
relations of Es are the result of the cosmological collapse of the
density fluctuations at the epoch when galactic scales became
non-linear, plus important modifications afterward due to early-time
merging in gas-rich systems. On the contrary, since the present day
non-linearity scale is that of galaxy clusters, and since the role of
gas dissipation is thought to be marginal in clusters life, the
cosmological collapse model and a mass-to-light ratio increasing with
cluster mass seem to be sufficient to account for their scaling
relations.

\section{Discussions and conclusions}
\label{sec:disc}
In this paper we have addressed the question of whether high mass DM halos
(that are thought to host the clusters of galaxies) do define the FP and the
other scaling relations observed for nearby clusters.  For that purpose, we
have analyzed a sample of 13 massive DM halos, obtained with high-resolution
N-body simulations in a $\Lambda$CDM cosmology.  After verifying that the DM
halos do follow the predictions of the spherical collapse model for
virialized systems, we have found that also their projected properties define
a FJ, a Kormendy and a FP--like relations.  This latter result is not trivial
and it can be traced back to the weak (structural and dynamical) homology
shown by the DM halos assembled via hierarchical cosmological merging.
However, the slopes of the DM halos scaling laws do not coincide with the
observed ones, and we have discussed what kind of systematic variations of
one or more of the structural and dynamical properties of the galaxy
distribution with cluster mass could account for such differences.  We have
shown that two of the three basic options can be discarded just by requiring
the simultaneous reproduction of the (edge-on) FP, FJ and Kormendy relations.
A solution that instead works remarkably well is to assume a cluster
mass-to-light ratio $\Upsilon$ increasing as a power law of the
luminosity. The required normalization and slope well agree with those
estimated observationally for real galaxy clusters. We have discussed two
possible causes for such a trend of $\Upsilon$, namely a systematic increase
of the {\it galaxy} mass-to-light ratio with cluster luminosity, or a
decrease of the baryonic mass fraction for increasing cluster total mass,
concluding that both the possibilities are consistent with the available
observational data.  In any case, it appears that {\it the FJ, Kormendy and
FP relations of nearby clusters of galaxies can be explained as the result of
the cosmological collapse of density fluctuations at the appropriate scales,
plus a systematic trend of the total mass-to-light ratio with the cluster
mass}.  Note that this is by no means a trivial result: in fact, it is known
that numerical simulations of galaxy dissipationless merging (where the
mass-to-light ratio is kept fixed) do reproduce well the edge-on FP, but
badly fail with the FJ and the Kormendy relations (NLC03b).

We next focused on the fact that Es follow a FJ relation with a
significantly different slope ($\sim 4$) compared to that of clusters
($\sim 2$), but show a similar trend of the mass-to-light ratio with
the system luminosity.  We have discussed the implications of this
observational evidences under the assumption that at the epoch of the
detachment from the Hubble flow, $M\propto\sigvir^3$ also at galactic
scales, as required by standard cosmology, and taking into account
several empirical and theoretical evidences suggesting that gas
dissipation and merging must have played an important role in galaxy
evolution.  Since dissipationless merging and baryonic dissipation
have competitive consequences on the system velocity dispersion, we
argue that {\it combined} effects of these two processes are required
to account for the slope of the FJ of elliptical galaxies.  We
therefore conclude that {\it the scaling relations of Es might be the
result of the cosmological collapse of density fluctuations at the
epoch when galactic scales became non-linear, plus successive
modifications due to (early-time) dissipative merging}.  This scenario
seems to be supported also by the similarity between the slopes of the
FJ and the $M_{\rm BH}$-$\sigma$ relations, and between the peak of
the star formation rate of the Universe and that of the quasar
luminosity function.

Before to conclude we point out that, while no observational data are
yet available for the scaling relations of galaxy clusters at high
redshift, the results that we have obtained allow us to speculate on
what could be expected.  On one hand, the predictions of the top-hat
model are the same at any redshift, and also weak homology of DM halos
in numerical simulations appears to hold at any epoch: thus, if the
dependence of the cluster $\Upsilon$ on $L$ changes accordingly to
passive evolution, scaling relations with the same slopes should be
found also at high redshift.  Even if with some uncertainties,
observational evidences for passive evolution of these cluster
properties already exist.  In fact, the characteristic luminosity of
the cluster LF increases with redshift consistently with pure, passive
luminosity evolution models (De Propris et al. 1999), and the same is
found for the stellar mass-to-light ratio $\Upsilon_*$ of elliptical
galaxies, as derived from the studies of their FP up to redshift
$z=1.27$ (e.g., van Dokkum \& Franx 1996; J{\o}rgensen et al 1999;
Kelson et al. 2000; Treu, M{\o}ller \& Bertin 2002; van Dokkum \&
Standford 2003).  On the other hand, a more complicate evolution of
$\Upsilon=\Upsilon_{\rm gal}\times M/M_{\rm gal}$ with $z$ might also
be expected, and thus the redshift variation of the FP zero-point and
slope could give important information on cluster and galaxy
evolution.  Although the fraction of Es in clusters appears not to
change with redshift, a substantial increase of the relative amount of
spirals with respect to S0 galaxies at higher $z$ is reported
(Dressler et al. 1997; Fasano et al. 2000).  In addition, also
$M/M_{\rm gal}$ might increase with $z$, if clusters form by
continuous accretion of galaxies from the field, in a DM potential
well that already settled on shorter time scales.  Of course, a
detailed prediction of the evolution of the cluster scaling relations
in this latter case is much more difficult.

Due to the interest of the questions addressed above, it is apparent
that with more available data (e.g., from the SDSS, RCS, MUNICS, 2dF,
2MASS surveys), a strong effort should be devoted to better determine
the cluster scaling relations while, on theoretical ground, a larger
set of simulated clusters, spanning a wider mass range and taking into
account also the presence of gas, should be performed and analyzed for
different choices of the cosmological parameters.

\acknowledgements {B.L. is grateful to Volker Springel, Simon White,
Naoki Yoshida and Gary Mamon for useful discussions and for their help
with the N-body simulations; Lauro Moscardini, Christophe Adami and
Massimo Meneghetti are also acknowledged for interesting
conversations. L.C. thanks Neta Bahcall and Jeremiah P. Ostriker, and
A.C. thanks Sophie Maurogordato for useful discussions. The anonymous
referee is acknowledged for useful comments that greatly improved the
paper. This work has been partially supported by the Italian Space
Agency grants ASI-I-R-105-00, ASI-I-R-037-01, and ASI-I-R-113-01, and
by the Italian Ministery (MIUR) grants COFIN2000 ``Halos and disks of
galaxies'', and COFIN2001 ``Clusters and groups of galaxies: the
interplay between dark and baryonic matter".}

\onecolumn

\end{document}